\documentclass[aps,prl,reprint,groupedaddress,amsmath,amssymb,showpacs]{revtex4-1}

\usepackage{graphicx}
\usepackage{natbib}

\begin{document}

\title{Energy Relaxation in the Integer Quantum Hall Regime}

\author{H. le Sueur}
\thanks{These authors contributed equally to this work.}
\affiliation{CNRS, Laboratoire de Photonique et de Nanostructures
(LPN) - Phynano team, route de Nozay, 91460 Marcoussis, France}
\author{C. Altimiras}
\thanks{These authors contributed equally to this work.}
\affiliation{CNRS, Laboratoire de Photonique et de Nanostructures
(LPN) - Phynano team, route de Nozay, 91460 Marcoussis, France}
\author{U. Gennser}
\affiliation{CNRS, Laboratoire de Photonique et de Nanostructures
(LPN) - Phynano team, route de Nozay, 91460 Marcoussis, France}
\author{A. Cavanna}
\affiliation{CNRS, Laboratoire de Photonique et de Nanostructures
(LPN) - Phynano team, route de Nozay, 91460 Marcoussis, France}
\author{D. Mailly}
\affiliation{CNRS, Laboratoire de Photonique et de Nanostructures
(LPN) - Phynano team, route de Nozay, 91460 Marcoussis, France}
\author{F. Pierre}
\email[Corresponding author: ]{frederic.pierre@lpn.cnrs.fr}
\affiliation{CNRS, Laboratoire de Photonique et de Nanostructures
(LPN) - Phynano team, route de Nozay, 91460 Marcoussis, France}

\date{\today}

\begin{abstract}
We investigate the energy exchanges along an electronic quantum channel realized in the integer quantum Hall regime at filling factor $\nu_L=2$. One of the two edge channels is driven out-of-equilibrium and the resulting electronic energy distribution is measured in the outer channel, after several propagation lengths $0.8~\mu$m$\leq L\leq30~\mu$m. Whereas there are no discernable energy transfers toward thermalized states, we find efficient energy redistribution between the two channels without particle exchanges. At long distances $L\geq10~\mu$m, the measured energy distribution is a hot Fermi function whose temperature is lower than expected for two interacting channels, which suggests the contribution of extra degrees of freedom. The observed short energy relaxation length challenges the usual description of quantum Hall excitations as quasiparticles localized in one edge channel.
\end{abstract}


\maketitle

The basic manifestation of the quantum Hall effect is a quantized Hall resistance $R_H=h/e^2\nu_L$, accompanied by a vanishing longitudinal resistance. In this regime, quantization of the two-dimensional cyclotron motion opens a large gap separating Landau levels in the bulk of the sample from the Fermi energy. The only available low energy excitations propagate along the edges, where the Landau levels cross the Fermi energy. The effective edge state theory suggests these excitations are prototypal one-dimensional chiral fermions (1DCF) \cite{halperin1982qhc,*buttiker1988abs}, each of the $\nu_L$ edge channels (EC) being identified with a one-dimensional conductor. Because back-scattering is forbidden by chirality, ECs are considered to be ideal ballistic quantum channels. Their similitude with light beams has inspired electronic analogues of quantum optics experiments \cite{ji2003emz,samuelsson2004twoeAB,olkhovskaya2008hom,neder2007i2e} and proposals for quantum information applications \cite{Ionicioiu2001qcwbe,*stace2004mowc}. However, the nature and decoherence of edge excitations are poorly understood, as highlighted by unexpected results obtained with electronic Mach-Zehnder interferometers: an unusual energy dependence of the interference fringes' visibility \cite{ji2003emz,bieri2009fbv}, a non-gaussian noise \cite{neder2007cdngsn} and a short coherence length \cite{litvin2007dsec,roulleau2008lphi}. Interactions between ECs and with their environment are seen as the key ingredient to explain these results (see e.g. \cite{seeling2001cfdephas,*chalker2007dimzi,*sukhorukov2007rdmzi,*youn2008noneqdephmzi,*degiovanni2009decrelqhec,levkivskyi2008dmzinu2}).

In the present experimental work, we investigate the interaction mechanisms taking place along an EC through the energy exchanges they induce. A similar approach was previously used on mesoscopic metal wires \cite{pothier1997relax} and on carbon nanotubes \cite{chen_noneq-cnt_2009}. Here we focus on the filling factor $\nu_L=2$, where two co-propagating ECs are present, and at which the above unexpected results were observed. Our experiment relies on the techniques we recently demonstrated to drive out-of-equilibrium an EC and to measure the resulting energy distribution $f(E)$ of 1DCF quasiparticles \cite{altimiras2009nesiqhr}. There, we drove out-of-equilibrium only the outer EC, and $f(E)$ was measured in the same EC after a short $0.8~\mu$m propagation distance, for which the energy redistribution is negligible. Here, we drive out-of-equilibrium selectively either the inner or the outer EC and probe $f(E)$ in the outer EC after various, much longer, propagation paths, up to $30~\mu$m. The electronic energy transfers, including those within and between the ECs, are revealed through changes in $f(E)$ along the edge. This gives us access to the underlying interaction mechanisms.

\begin{figure}[!t]
\includegraphics[width=1\columnwidth]{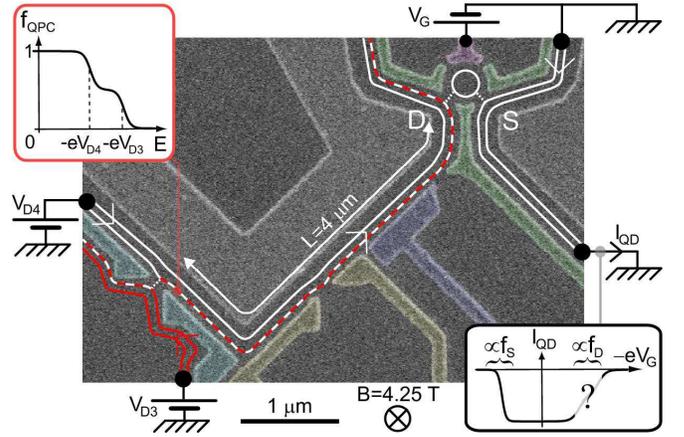}
\caption{(Color online) Sample micrograph: metallic gates appear bright; the two widest gates (not colorized) are grounded. The current propagates counter clockwise along two edge channels (EC) depicted by lines. White dashed lines indicate intermediate EC transmissions. At the output of the voltage biased quantum point contact (left in figure), the electronic energy distribution is a double step (left inset) in the partly transmitted EC (dashed outer EC in figure). After an adjustable propagation distance ($L=4~\mu$m in figure), the energy distribution $f_D$ in the outer EC is measured using a quantum dot (white circle, see right inset).}
\label{Fig1}
\end{figure}
The measured sample displayed in Fig.~1 was tailored in a two-dimensional electron gas realized in a GaAs/Ga(Al)As heterojunction of density $2~10^{15}~\mathrm{m}^{-2}$, mobility $\mu=250~\mathrm{m}^2\mathrm{V}^{-1}\mathrm{s}^{-1}$, and measured in a dilution refrigerator of base temperature 30~mK \cite{si}. The relevant ECs are defined by voltage biased surface metallic gates (except a small portion defined by mesa etching for the longest propagation paths, see \cite{si}). The energy distribution $f_D(E)$ in the outer EC at the drain (D) side of a quantum dot (QD, white circle in Fig.~1) is probed using the QD as an energy spectrometer, as has already been described in \cite{altimiras2009nesiqhr}: We record the differential conductance $\partial I_{\mathrm{QD}}(V_G)/\partial V_{G} \propto \partial f_D(E)/\partial E$, with $I_{\mathrm{QD}}$ the tunnel current across the small QD set to have a single active electronic level, while sweeping the voltage $V_G$ applied to a capacitively coupled gate \cite{altimiras2009nesiqhr}. The path length $L\in\{0.8,2.2,4,10,30\}~\mu$m is tuned in-situ by first choosing the pair of metallic gates that define the quantum point contact (QPC) at which a non-equilibrium energy distribution $f_\mathrm{QPC}$ is induced, and then by applying a negative voltage to selected gates to define the path between the QPC and the QD. A non-equilibrium smeared double step $f_\mathrm{QPC}(E)$ \cite{altimiras2009nesiqhr} is induced at the output of the voltage biased QPC selectively in the outer or inner EC by adjusting the QPC's conductance to $0.5~e^2/h$ or $1.5~e^2/h$, which are illustrated in Fig.~2(a) and 3(a), respectively.

\begin{figure}[!t]
\includegraphics[width=2.75in]{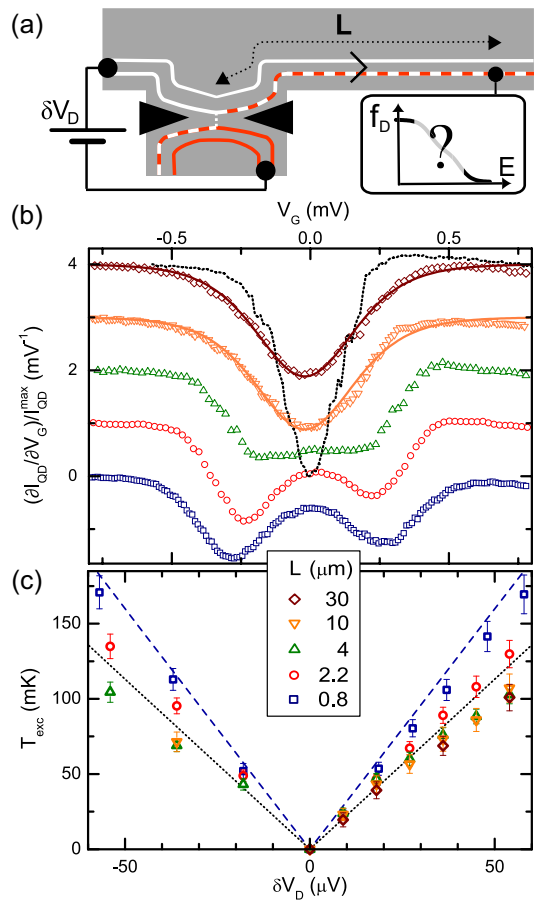}
\centering
\caption{(Color online) (a) The \textit{outer} EC is driven out-of-equilibrium. (b) Raw data (symbols) at $\delta V_D=36~\mu$V, shifted vertically for several $L$. The non-equilibrium double dip relaxes over $L_{inel} \approx 3~\mu$m toward a dip broader than the equilibrium dip at $\delta V_D=0$ (dotted line). Solid lines are calculations with a Fermi distribution at $85~\mathrm{mK}$. (c) Excess temperatures extracted from the data (symbols) and prediction at the QPC output (dashed line). The outer EC cools down as $L$ is increased and saturates at a value below expectations for two interacting ECs (dotted line, see text).}
\label{Fig2}
\end{figure}
First, we generate a non-equilibrium energy distribution in the measured \textit{outer} EC (Fig.~2(a)). The raw $\partial I_{\mathrm{QD}}/\partial V_{G}$ data are shown in Fig.~2(b) for several lengths $L$, at fixed QPC voltage bias $\delta V_D=36~\mu$V. For the shortest propagation length $L=0.8~\mu$m, we find a double dip close to expectations for non-interacting ECs and, consequently, that energy exchanges are small on this scale \cite{altimiras2009nesiqhr}. As $L$ is increased the signal evolves toward a single dip. This demonstrates energy exchanges, which occur here on a characteristic length $L_{inel} \approx 3~\mu$m \cite{si}. At the two longest propagation paths, we find the \textit{same} broad dip within experimental accuracy. It corresponds to a drain Fermi distribution of temperature $T_{\mathrm{hot}}=85~$mK (solid lines on top of data at $L=10$ and $30~\mu$m), much larger than the equilibrium dip's temperature $T_{\mathrm{eq}}=40~$mK (data at $\delta V_D=0$ are shown for comparison as a dotted line). Complementary tests were performed to ascertain the observed energy exchanges are not artifacts \cite{si}.

We now investigate the interaction mechanisms responsible for the established energy exchanges. A simple mechanism could be the tunneling of charges between co-propagating ECs, but we found it is here negligible \cite{si}. In particular, anomalous quantum Hall effect measurements \cite{vanwees1989aiqhe} showed that there is here no equilibration along the considered paths between the different electrochemical potentials of the two co-propagating ECs. Important information to elucidate the interaction mechanisms can be obtained from the total energy $E_\mathrm{out}$ of the probed outer EC's 1DCFs. Let us consider several scenarios. If \textit{(i)} interactions are essentially between 1DCFs in the same EC, then energy conservation in the stationary regime implies $E_\mathrm{out}$ is unchanged along the propagation path. Else, if \textit{(ii)} there is a significant interaction with thermalized states, such as the many quasiparticles within the surface metal gates along sample edges or the phonons, then $E_\mathrm{out}$ should relax toward its cold equilibrium value $E_\mathrm{out}(\delta V_D=0)$; or, if either the coupling constant or the density of these states vanishes at low energies, toward a fixed value at large $\delta V_D$. Last \textit{(iii)}, if interactions are essentially with other co-propagating states, then the injected energy redistributes. Therefore $E_\mathrm{out}$ should decrease to a value above $E_\mathrm{out}(\delta V_D=0)$ by an amount proportional to the injected energy. The co-propagating states could be the inner EC's 1DCFs or/and additional internal EC modes \cite{aleiner1994nee,*chamon1994er} that are predicted to exist in most situations due to edge reconstruction \cite{chklovskii1992eec,*macdonald1993er}.

Figure 2(c) shows the outer EC's energy for various $L$ and $\delta V_D$ as the generalized excess temperature $T_\mathrm{exc}\equiv \sqrt{6 (E_\mathrm{out}-E_\mathrm{out}(\delta V_D=0))/\nu\pi^2k_B^2}$ (symbols), with $\nu$ the outer EC's density of states per unit length and energy. The ratio  $E_\mathrm{out}/\nu$ can be obtained from $f_D$ using
\begin{equation}
E_\mathrm{out}/\nu= \int (E-\mu)(f_D(E)-\theta(\mu-E))dE,\label{EsNu}
\end{equation}
with $\theta(E)$ the step function and $\mu$ the electrochemical potential (the full procedure to extract $T_\mathrm{exc}$ is detailed in \cite{si}). We find $T_\mathrm{exc}$ relaxes as $L$ increases, from a value very close to the QPC output prediction $T_\mathrm{exc}^\mathrm{qpc}=\sqrt{3} e | \delta V_D| /(2\pi k_B)$ (dashed line) at $L=0.8~\mu$m, down to $T_\mathrm{exc}\approx (0.61\pm0.04) \times T_\mathrm{exc}^\mathrm{qpc}$ at $L=10$ and $30~\mu$m for $|\delta V_D|>20~\mu$eV. The saturation of $T_\mathrm{exc}$ at long propagation lengths to a value proportional to $T_\mathrm{exc}^\mathrm{qpc}$ at the QPC output is \textit{incompatible} with significant dissipation toward thermalized states on the probed length scales (scenario \textit{ii}). Instead, this observation corresponds to expectations for interactions with co-propagating states (scenario \textit{iii}). Last, energy exchanges between 1DCFs of the same EC (scenario \textit{i}) are \textit{relatively weak} compared to the dominant mechanism. Indeed, they preserve $T_\mathrm{exc}$, whereas we find $f_D$ and $T_\mathrm{exc}$ evolve on the same length scale, as seen in Fig.~2(b) \& (c). Additional experiments not shown here further demonstrate that this energy exchange mechanism is negligible for $L\leq10~\mu$m \cite{altimiras2010ctrlrelaxqhr}.

\begin{figure}[!t]
\includegraphics[width=2.75in]{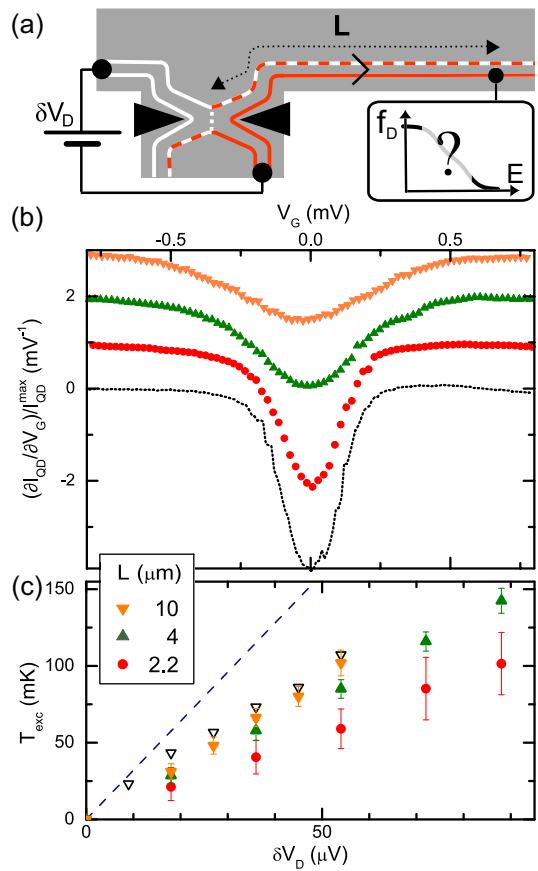}
\centering
\caption{(Color online) (a) The \textit{inner} EC is driven out-of-equilibrium. (b) Raw data at $\delta V_D=0$ (dotted line) and $\delta V_D=54~\mu$V (symbols), shifted vertically for several $L$. The dip broadens as $L$ is increased. (c) Excess temperatures extracted from the data (full symbols) and prediction at the QPC output (dashed line). The outer EC heats up as $L$ is increased, up to an excess temperature close to that when driving the outer EC out-of-equilibrium ($T_\mathrm{exc}(L=10~\mu\mathrm{m})$ in Fig.~2(c) are shown here as open symbols ($\triangledown$)).}
\label{Fig3}
\end{figure}

The data are compatible with energy redistribution with co-propagating states, but which states? It is most natural to assume the 1DCFs of the two co-propagating ECs exchange energy. This hypothesis can be tested directly by generating a non-equilibrium energy distribution in the \textit{inner} EC ($G_{\mathrm{QPC}} \simeq 1.5 e^2/h$), with $f_D$ still being measured in the \textit{outer} EC (see Fig.~3(a)). Figure~3(b) shows raw data obtained in this configuration at $\delta V_D=54~\mu$V for several $L$. We find that the dip broadens as $L$ is increased, and therefore that the outer EC heats up. This unambiguously demonstrates energy exchanges between ECs. Figure~3(c) shows $T_\mathrm{exc}$ in the outer EC (symbols), which increases with $L$ as expected from the raw data. Note that $T_\mathrm{exc}(L=10~\mu$m) is approximately independent of which of the inner or the outer EC is driven out-of-equilibrium, as would be expected for a complete energy current equipartition between ECs.

These results are in qualitative agreement with recent investigations of dephasing at $\nu_L=2$, which established current noise in one EC reduces phase coherence in the second EC \cite{roulleau2008noisedephasing,neder2007cdngsn}. The dephasing length $L_\phi(T) \simeq 20~\mu \mathrm{m}/(T/20~\mathrm{mK})$ \cite{roulleau2008lphi} can be compared to the inelastic length. Using the injected excess temperature $T_\mathrm{exc}^\mathrm{qpc}= 115~\mathrm{mK}$ at $\delta V_D=36~\mu\mathrm{V}$, we find $L_\phi(115~\mathrm{mK}) \simeq 3.5~\mu$m, similar to the corresponding $L_\mathrm{inel}= 2.5\pm0.4~\mu$m \cite{si}. This strengthens the case for a same physical mechanism at the root of both dephasing and energy exchanges. However, contrary to dephasing \cite{roulleau2008noisedephasing}, energy exchanges cannot be accounted for by low frequency noise within perturbation theories.

We now discuss different theoretical models aiming at explaining the present data. Within the widespread picture of 1DCF quasiparticles, the minimal approach is to include interactions between co-propagating ECs as a small perturbation. However, in absence of disorder, energy exchanges between 1DCFs of different drift velocities $v_D$ would be essentially suppressed, due to combined energy and momentum conservations. Therefore, it is crucial to assume a sufficient disorder to break momentum conservation. Motivated by the present work, Lunde \textit{et al.} modeled inter EC interactions as a density-density coupling, where disorder changes the coupling coefficient along the edge with a correlation length $\ell$ \cite{lunde_interaction_2010}. Within this model, $T_\mathrm{exc}(L/L_0,\delta V_D)$ was obtained up to an unknown length scaling factor $L_0$. Comparing with the data, it was found that the non-linear shape of $T_\mathrm{exc}(\delta V_D)$ can be reproduced using a reasonable micron-scale $\ell$ \cite{lunde_interaction_2010}. On the other hand, general arguments imply that two weakly interacting 1DCF branches cannot result in $T_\mathrm{exc}<T_\mathrm{exc}^\mathrm{qpc}/\sqrt{2}$ at saturation \cite{si,lunde_interaction_2010}. Surprisingly, we find $T_\mathrm{exc}$ at long $L$ saturates about $\approx 13\%$ \textit{below} this lower bound (displayed as a dotted line in Fig.~2(b)). Such discrepancy is significantly larger than experimental error bars. Although a good agreement data-theory was reached in \cite{lunde_interaction_2010} assuming \textit{ad-hoc} the presence of a hidden third EC, one may wonder if the discrepancy results from the perturbative treatment of interactions. Note that the weak interaction hypothesis could not be checked in \cite{lunde_interaction_2010}, due to the unknown length scaling factor in the theory.

Alternatively, density-density interactions between co-propagating 1DCFs can be handled non-perturbatively using the bosonization technique \cite{giamarchi2003qp1d}. Within this framework, edge states are depicted as collective magnetoplasmon modes. For strong enough interactions, these are fully delocalized over the ECs \cite{wen1990epgeefqs,levkivskyi2008dmzinu2}. At filling factor 2, where the two ECs have opposite spin polarities, this yields spinless charge waves and chargeless spin waves propagating at different velocities. These edge states appear strikingly different from quasiparticles, where both charge and spin propagate at the same speed. Motivated by the present experiment, $T_\mathrm{exc}$ was recently calculated in the bosonization framework \cite{degiovanni_plasmon_2010}. Assuming strong interactions and a standard drift velocity $10^5~\mathrm{m}/$s, calculations are found to reproduce the measured non-linear shape of $T_\mathrm{exc}(\delta V_D)$ and also the energy relaxation length scale, without the need to introduce disorder \cite{degiovanni_plasmon_2010}. However, the same lower bound $T_\mathrm{exc}^\mathrm{qpc}/\sqrt{2}$ was confirmed for arbitrary interaction strength between two 1DCF branches. In \cite{degiovanni_plasmon_2010}, the data are reproduced quantitatively by assuming \textit{ad-hoc} $25\%$ of the energy leaks out toward other degrees of freedom.

The main outcome of the data-theory comparisons is that additional states need to be taken into account. Experimental observations, in particular the saturation at the same hot Fermi distribution for both $L=10$ and $30~\mu$m, put stringent constraints on these states. The predicted internal EC modes mentioned in scenario \textit{(iii)} seem plausible candidates. However, additional experiments not shown here demonstrate that 1DCFs and internal modes localized in the same outer EC do not exchange energy \cite{altimiras2010ctrlrelaxqhr}. Although this weakens the internal modes hypothesis, note that energy exchanges with the inner EC's internal modes were not dismissed.

One conceptually important question concerns the nature of the pertinent edge excitations. Are these better described as Fermi quasiparticles localized in an EC or as delocalized bosonic collective states? The above comparison with theories did not permit to discriminate. Nevertheless, the experimental results can be used to test whether the quasiparticle description is self-consistent. Indeed, a lower bound for the 1DCF's lifetime can be obtained from $L_\mathrm{inel}$, by using the range of drift velocities $v_D\in [0.5,5]~10^5$~m/s measured in similar structures at $\nu_L=2$ \cite{si,kamata_vdQHR_2010}. Applying the time-energy uncertainty relation, one finds for $\delta V_D=36~\mu \mathrm{V}$ that the energy linewidth of 1DCF states $\Delta E> \hbar v_D / 2L_\mathrm{inel}\in [6,70]~\mu$eV is of the same order or larger than their characteristic energy $k_B T_\mathrm{exc}^\mathrm{qpc}(\delta V_D=36~\mu \mathrm{V})\approx 10~\mu$eV, and therefore are ill-defined electronic edge excitations. Consequently, although the 1DCF representation of edge states is very powerful at short distances, the observed short $L_\mathrm{inel}$ challenges the description of quantum Hall excitations as quasiparticles localized in one edge channel.

\begin{acknowledgments}
The authors gratefully acknowledge discussions with A.\ Anthore, M.\ B\"{u}ttiker, P.\ Degiovanni, C.\ Glattli, P.\ Joyez, I.\ Neder, F.\ Portier, H.\ Pothier, P.\ Roche, E.\ Sukhorukov. This work was supported by the ANR (ANR-05-NANO-028-03).
\end{acknowledgments}

%

\end{document}


\title{Supplementary Material for \\`Energy Relaxation in the Integer Quantum Hall Regime'}

\author{H. le Sueur}
\thanks{These authors contributed equally to this work.}
\affiliation{CNRS, Laboratoire de Photonique et de Nanostructures
(LPN) - Phynano team, route de Nozay, 91460 Marcoussis, France}
\author{C. Altimiras}
\thanks{These authors contributed equally to this work.}
\affiliation{CNRS, Laboratoire de Photonique et de Nanostructures
(LPN) - Phynano team, route de Nozay, 91460 Marcoussis, France}
\author{U. Gennser}
\affiliation{CNRS, Laboratoire de Photonique et de Nanostructures
(LPN) - Phynano team, route de Nozay, 91460 Marcoussis, France}
\author{A. Cavanna}
\affiliation{CNRS, Laboratoire de Photonique et de Nanostructures
(LPN) - Phynano team, route de Nozay, 91460 Marcoussis, France}
\author{D. Mailly}
\affiliation{CNRS, Laboratoire de Photonique et de Nanostructures
(LPN) - Phynano team, route de Nozay, 91460 Marcoussis, France}
\author{F. Pierre}
\email[Corresponding author: ]{frederic.pierre@lpn.cnrs.fr}
\affiliation{CNRS, Laboratoire de Photonique et de Nanostructures
(LPN) - Phynano team, route de Nozay, 91460 Marcoussis, France}

\maketitle

\section{Methods}

\subsection{Sample fabrication}

The sample was realized in a standard GaAs/Ga(Al)As two dimensional electron gas located 105~nm below the surface, of density $2~10^{15}~\mathrm{m}^{-2}$, Fermi energy $80~$K and mobility $250~\mathrm{m}^2V^{-1}s^{-1}$. Note that the same GaAs/Ga(Al)As heterojunction was used formerly to perform the Mach-Zehnder experiments with edge states reported in \cite{roulleau2008lphi,roulleau2008noisedephasing}, and to demonstrate the non-equilibrium edge channel spectroscopy \cite{altimiras2009nesiqhr}. The silicon (dopant) concentration was adjusted to optimize the Hall resistance quantization. The sample was patterned using e-beam lithography followed by chemical etching of the heterojunction and by deposition of metallic gates at the surface.

\subsection{Experimental techniques}

Conductance measurements were performed in a dilution refrigerator of base temperature 30~mK. All measurement lines were filtered by commercial $\pi$-filters on top of the cryostat. We also carefully filtered and thermalized them at the low temperature stages, by using 1~m long resistive twisted pairs ($300~\Omega /$m) inserted inside 260~$\mu$m inner diameter CuNi tubes, tightly wrapped around a copper plate screwed to the mixing chamber. The sample was further protected from spurious high energy photons by two shields, both at base temperature.

The sample was current biased using a voltage source in series with a $10~$M$\Omega$ or $100~$M$\Omega$ polarization resistance at room temperature. Taking advantage of the well-defined quantum Hall resistance (12.906~k$\Omega$), currents across the sample were converted on-chip into voltages and measured with low noise room temperature voltage amplifiers. To limit artifacts due to slowly moving charges nearby the quantum dot (QD), we systematically measured several successive gate voltage sweeps $I_{\mathrm{QD}}(V_G)$, checked that the data fall on top of each other, and verified that the sum rule $\int (\partial I_{\mathrm{QD}}/\partial V_G) dV_G \simeq 0$ is obeyed. To avoid artificial heating, AC voltages were always kept smaller than $k_BT/e$.

\subsection{Energy distribution spectroscopy with a quantum dot}

We measured the energy distribution $f_D$ in the outer EC using a QD as a tunable energy filter \cite{altimiras2009nesiqhr}. Assuming a single active QD level of energy $E_{lev}$, constant tunneling density of states and tunnel rates, the QD current $I_{\mathrm{QD}}$ in the sequential tunnel regime reads \cite{kouwenhoven1997etqd}
\begin{equation}
I_{\mathrm{QD}}=I_{\mathrm{QD}}^{max}(f_S(E_{lev})-f_D(E_{lev})), \label{SIIdot}
\end{equation}
where $I_{\mathrm{QD}}^{max}$ is the maximum QD current and $D$ ($S$) refers to the drain (source) outer EC located on the left (right) side of the QD. We extracted $f_S$ and $f_D$ separately by applying a sufficiently large source-drain voltage (right inset in Fig.~1). In practice we used $V_{D4}=-88~\mu$V (see Supplementary Figure~1). The probed energy $E_{lev}=E_0-e\eta_G V_G$ was swept using $V_G$, with $E_0$ an unimportant offset and $\eta_G$ the gate voltage-to-energy lever arm calibrated by temperature and non-linear QD characterizations (see section `Experimentally determined lever arm', below). Raw data $\partial I_{\mathrm{QD}}(V_G)/\partial V_G$ measured by lock-in techniques are proportional to $\partial f_{D,S} (E)/\partial E$. It is useful to remember that in the present setup a Fermi distribution in the drain (source) edge channel appears as a single negative dip (positive peak) in $\partial I_{\mathrm{QD}}(V_G)/\partial V_G$, whose width and inverse amplitude are proportional to temperature.

As pointed out above, the simple expression of Supplementary Equation~\ref{SIIdot} assumes only one QD level contributes to $I_{\mathrm{QD}}$, and neglects the energy dependence of the electrodes tunneling density of states and of the tunnel rates in and out the QD. In practice, the validity of these hypotheses were checked with a standard non-linear QD characterization \cite{kouwenhoven1997etqd}, and by comparing mixing chamber temperatures with fit temperatures obtained within this framework (see Figure~2 in \cite{altimiras2009nesiqhr}).

Importantly, the measured energy distribution is that of 1DCF quasiparticle excitations in the probed edge channel. In particular, the tunnel coupled QD does not probe the predicted additional edge excitations \cite{aleiner1994nee} corresponding to transverse charge oscillations across the finite width of edge channels \cite{chklovskii1992eec} (see \cite{altimiras2009nesiqhr} and references therein for a detailed discussion).

\subsection{Practical realization of the different experimental configurations}

The experimental configuration was tuned in-situ with the bias voltages applied to surface metallic gates. Supplementary Figure~1 shows how the various propagation lengths $L\in \{0.8,2.2,4,10,30\}~\mu$m were realized in practice. Note that the data shown here were obtained in the same cooldown.

\begin{figure*}[!ht]
\center
\includegraphics[width=6.3in]{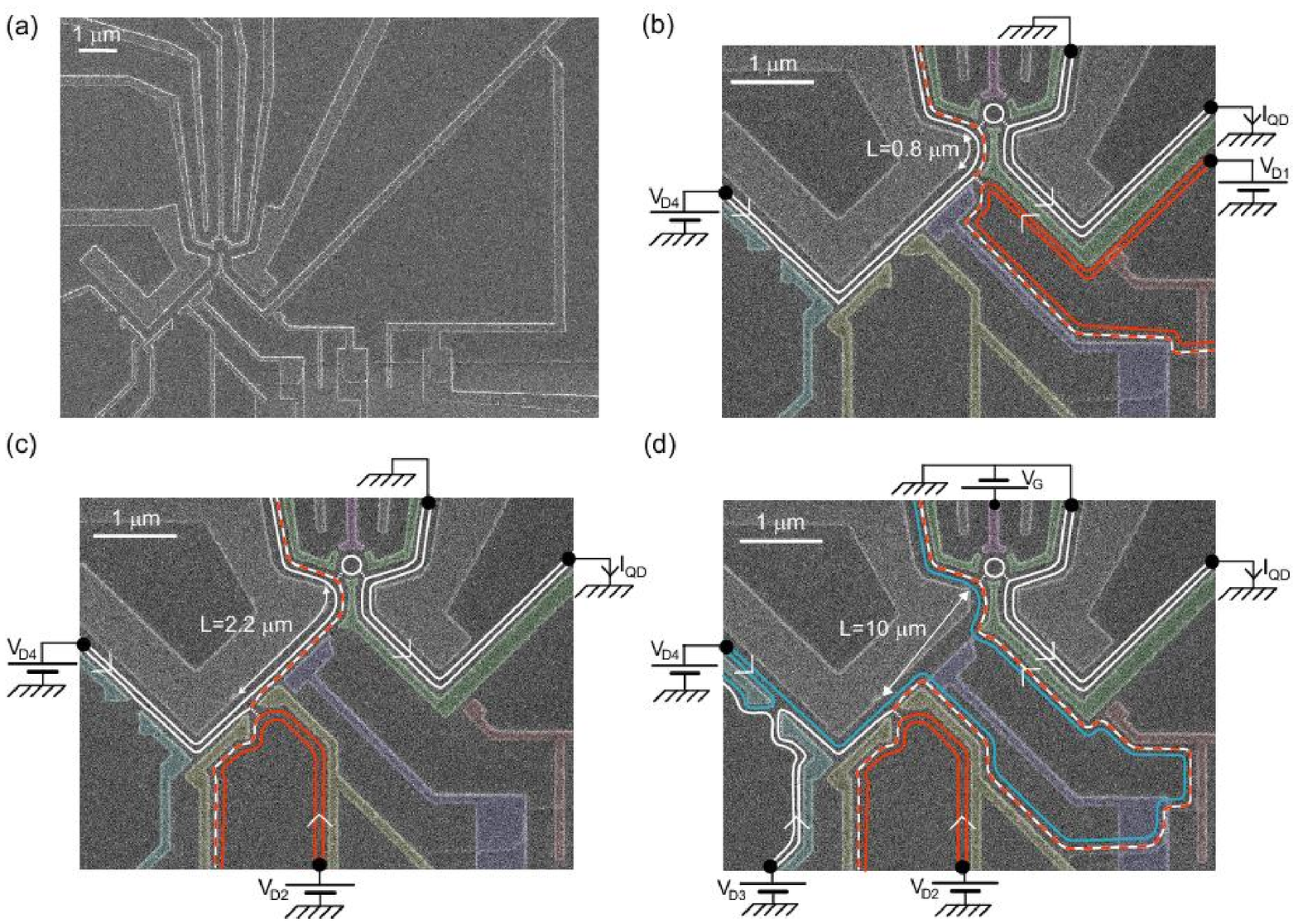}
\caption{(a), e-beam micrograph of the sample. (b), (c) \& (d), Experimental realizations of propagation lengths $L=0.8~\mu$m, $2.2~\mu$m and $10~\mu$m, respectively. The realization of length $L=4~\mu$m is shown in Figure~1. The length $L=30~\mu m$ uses the rightmost gate in (a) to return the edge channels toward the QD, with the same injection QPC (yellow split gate) as $L=2.2~\mu$m and $10~\mu$m. Here, the outer EC is shown half transmitted and the inner EC fully reflected ($G_{QPC}=0.5e^2/h$). At $G_{QPC}=1.5e^2/h$ the outer EC is fully transmitted and the inner EC half transmitted, in order to inject energy in the latter. In (d), we also show how the two ECs upstream of the partitioning QPC are biased at different voltages. This allowed us to test the absence of charge tunneling between the co-propagating ECs (see `Experimental test of charge tunneling between the co-propagating edge channels' in `Methods').}
\label{SIfig1}
\end{figure*}

\subsection{Experimentally determined lever arm}

The QD gate voltage $V_G$ to energy $E$ lever arm $\eta_G$ sets the energy scale of the energy distribution functions, and appears as a scaling factor in the extracted excess temperatures shown in Figures~2 and 3.

We used three methods to extract this parameter.
We first performed a characterization of the QD in the non-linear regime \cite{kouwenhoven1997etqd} and extracted $\eta_G$ from the slopes of the Coulomb diamond $\partial I_{\mathrm{QD}}/\partial V_G(V_G,V_D)$ (hereafter called method~1).
We also measured $\partial I_{\mathrm{QD}}/\partial V_G(V_G,V_D=-88~\mu\mathrm{V})$ at several temperatures and extracted $\eta_G$ from the scaling between the measured mixing chamber temperatures and the fit temperatures $T_S$ ($T_D$) of the source peak (drain dip) obtained assuming Fermi functions (hereafter called method~2 (method~3)) (see Figure~2 in \cite{altimiras2009nesiqhr}).

We defined $\eta_G$ as the average of the three values obtained with these methods, and used the corresponding standard error to define uncertainty on $\eta_G$. Note that a different procedure was used in ref.~\cite{altimiras2009nesiqhr}, where uncertainty was evaluated by finding the range of $\eta_G$ that account for most values of $T_{S,D}$ at temperatures above $50~$mK.

A full  QD calibration (method 1 to 3) was performed each time we changed a surface gate nearby the QD by an important amount. Indeed, we found that capacitive cross-talks between the QD and the surface metallic gates used to manipulate the ECs' paths were not negligible. The average values of $\eta_G$ used in the present work are recapitulated in Table~\ref{SItabEtaG} with their relative standard errors. Note that we found small variations between the various average values of $\eta_G$, well within standard errors. This suggests that the relative error bars between experimental configurations is smaller than the absolute error bar displayed in all figures.

\begin{table}
\center
\begin{tabular}{|c|c|c|c|c|c|}
  \hline
$\{L,G_{QPC}\}$ & $\eta_G$ & $\Delta \eta_G/\eta_G$  \\
$\{(\mu\mathrm{m}),(e^2/h)\}$ &   & $(\%)$  \\
\hline
$\{0.8,0.5\}^+$ & 0.0593 & 5.3 \\
$\{4,1.5\}^+$ & 0.0610 & 3.5 \\
$\{4,0.5\}$ & 0.0610 & 3.5 \\
$\{2.2,0.5\}$ & 0.0610 & 3.5 \\
$\{2.2,1.5\}$ & 0.0610 & 3.5 \\
$\{10,0.5\}^+$ & 0.0606 & 5.7 \\
$\{30,0.5\}$ & 0.0606 & 5.7 \\
$\{10,1.5\}^+$ & 0.0598 & 6.3 \\
\hline
\end{tabular}
\caption{Used lever arms $\eta_G$ and their relative standard errors $\Delta \eta_G/\eta_G$. The symbol ($^+$) points out experimental configurations, characterized by $\{L,G_{QPC}\}$, for which the full QD calibration was performed. }
\label{SItabEtaG}
\end{table}

\subsection{Experimental test of charge tunneling between the co-propagating edge channels}

It is known that the electrochemical potentials of co-propagating ECs equilibrate on large propagation distances \cite{vanwees1989aiqhe,komiyama1989viqhe,alphenaar1990seiqhe}. At low temperatures and filling factor 2, macroscopic equilibration lengths of 1~mm were reported \cite{muller1992eleiqhe}.

First, the most straightforward approach to test the presence of tunneling is to perform an anomalous quantum Hall effect measurement \cite{vanwees1989aiqhe}, namely biasing the two edge channels at different potentials and measuring them separately after some propagation distance. This can be done by using injection and measurement QPCs set to the conductance $G_{QPC}=e^2/h$ to separate the two ECs. We did this measurement, and also a variation of this measurement in which the injection QPC was set to $G_{QPC}\simeq 0.5 e^2/h$, for $L=30~\mu$m and for the largest applied bias voltage $\delta V_D=\pm 54~\mu$V. We found, at base temperature $T=30~$mK, that tunneling between co-propagating edge channels was always negligible (less than 1\% of the population difference).

Second, in order to perform this test in the non-equilibrium situation investigated, simultaneously to data acquisition, we made use of the fact that in absence of charge tunneling between ECs, the inner EC is only capacitively coupled to the QD, as is the plunger gate. Therefore, the presence of tunneling between co-propagating ECs shows up as deviations from a strict proportionality between $\partial I_{\mathrm{QD}}/\partial V_{in}(V_G)$ and $\partial I_{\mathrm{QD}}/\partial V_{G}(V_G)$. We systematically applied an AC modulation $e\delta V_{in}$ to the inner EC electrochemical potential at a specific frequency, different from the outer EC's modulation frequencies. We could do this by separating the two ECs upstream of the injection QPC (see Figure~1 with $V_{in}=V_{D4}$), except for $L=4~\mu$m, where the injection QPC is the foremost upstream. We then measured $\partial I_{\mathrm{QD}}/\partial V_{in}$ by lock-in techniques, and checked that there were no such deviations at the largest applied bias voltage in each experimental configurations.
Note that we verified the pertinence of this second test at a larger fridge temperature $T=190~$mK, where the first test (i.e. anomalous quantum Hall effect measurements) showed the presence of tunneling between co-propagating ECs ($8.7\%$, $5.0\%$ and $2.3\%$ equilibration at $\delta V_D=54~\mu$V, $36~\mu$V and 0~V, respectively, after a propagation distance $L=30~\mu$m). The fact that we also observed significant deviations from $\partial I_{\mathrm{QD}}/\partial V_{in} \propto \partial I_{\mathrm{QD}}/\partial V_{G}$ demonstrates the pertinence of this second test.

\subsection{Procedure to extract $T_\mathrm{exc}$ and the displayed error bars from the data}

Following \cite{altimiras2009nesiqhr}, the displayed generalized excess temperature $T_\mathrm{exc}\equiv \sqrt{6 (E_\mathrm{out}-E_\mathrm{out}(\delta V_D=0))/\nu\pi^2k_B^2}$, with $\nu$ the outer EC density of states per unit length and energy, is obtained from the energy distribution $f_D$ using
\begin{equation}
E_\mathrm{out}/\nu= \int (E-\mu)(f_D(E)-\theta(\mu-E))dE,\label{SIEsNu}
\end{equation}
with $\theta(E)$ the step function, and $\mu$ the electrochemical potential given by
\begin{equation}
\mu=E_{min}+\int_{E_{min}}^{E_{max}}f_D(E)dE, \label{SImu}
\end{equation}
with $E_{min}$ ($E_{max}$) an energy under (above) which we assumed $f(E)$ remains 1 ($0$).

However, in order to limit artifacts related to finite signal-to-noise ratio, to the finite energy window probed and to the simple QD model used, we also always tried to fit the data assuming that $f_D$ is the weighted sum of two Fermi functions. This yielded an alternative value for $T_\mathrm{exc}$. First, at equilibrium, the Fermi fit gave an estimate of deviations of our detector from the simple QD model \cite{altimiras2009nesiqhr}. Second, as long as the accuracy of the non-equilibrium fits was found equal to or better than that of the reference at equilibrium (most often the case), $T_\mathrm{exc}$ was taken as the average value of the fit and the integral~(\ref{SIEsNu}) procedures.

Displayed error bars on $T_\mathrm{exc}$ include the two independent contributions of the standard deviation between extraction procedures for $T_\mathrm{exc}$ and for $\eta_G$. In practice, we found in most cases that the error is dominated by the latter contribution.

\subsection{Procedure to extract $L_{inel}$ from the data}

The inelastic length $L_{inel}$ is the length scale for energy exchanges in a given non-equilibrium situation. In practice the non-equilibrium situation is here characterized by the voltage $\delta V_D$ applied to a half transmitting QPC. We extract the quantity $L_{inel}(\delta V_D)$ by fitting $T_\mathrm{exc}(L)$ at a fixed $\delta V_D$ with the exponential function
\begin{equation}
T_\mathrm{exc}^{fit}(L) \equiv \\ (T_\mathrm{exc}^\mathrm{qpc}-T_\mathrm{exc}^\mathrm{sat})\exp{(-L/L_{inel})}+T_\mathrm{exc}^\mathrm{sat},
\end{equation}
where $T_\mathrm{exc}^\mathrm{qpc}=\sqrt{3} e | \delta V_D| /(2\pi k_B)$ and $T_\mathrm{exc}^\mathrm{sat}$ is a second fit parameter that corresponds to the excess temperature at large Ls.

The values of $L_{inel}(\delta V_D)$ extracted by this procedure are recapitulated in Table~II. Error bars $\Delta L$ on $L_{inel}$ are standard errors obtained taking into account error bars in the extracted $T_\mathrm{exc}$. $T_\mathrm{exc}^{fit}(L)$ corresponding to $L_{inel}$ shown in Table~II are displayed as continuous lines in Supplementary Fig.~2.
\begin{figure}[!ht]
\includegraphics[width=\columnwidth]{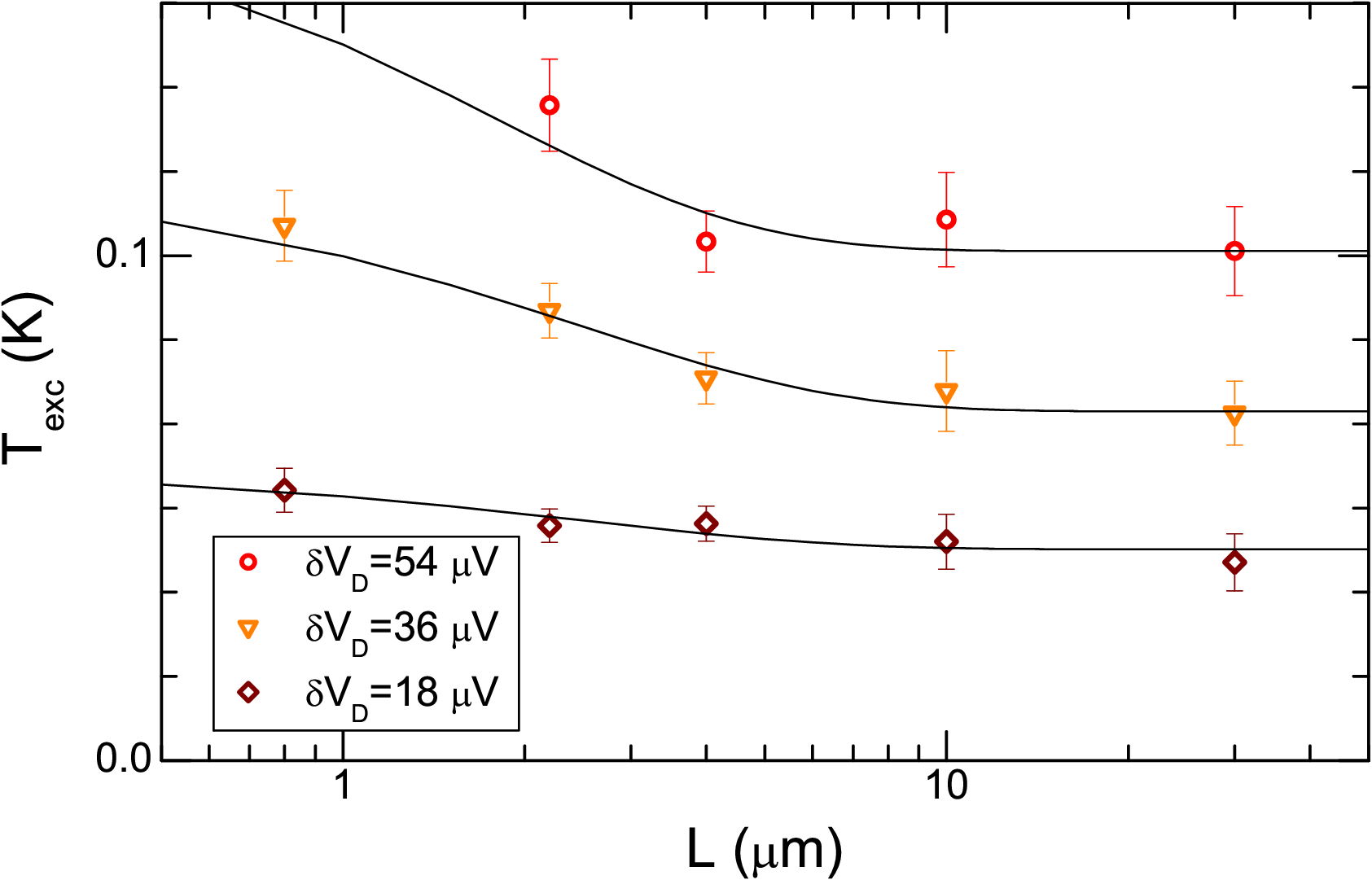}
\caption{Fits of $T_\mathrm{exc}(L)$ to extract $L_{inel}$ shown in Table~II.}
\label{SIFigLinel}
\end{figure}

\begin{table}
\center
\begin{tabular}{|c|c|c|c|c|c|}
  \hline
$\delta V_D$ & $L_{inel}$ & $\Delta L$ \\
$(\mu\mathrm{V})$ & $(\mu\mathrm{m})$  & $(\mu\mathrm{m})$ \\
\hline
$18$ & 2.5 & 0.85 \\
$36$ & 2.5 & 0.4 \\
$54$ & 1.8 & 0.65 \\
\hline
\end{tabular}
\caption{$L_{inel}$ and corresponding standard error obtained by fitting $T_\mathrm{exc}(L)$ (see text).}
\label{SItabLin}
\end{table}

\section{Supplementary data}

We here show data measured after the shortest propagation length $L=0.8~\mu$m and for an injection QPC set to the conductance $G_{QPC}\simeq0.5e^2/h$, in order to illustrate the tuning out-of-equilibrium and the spectroscopy of the energy distribution function (see Supplementary Figure~3).
Along this short propagation distance, we find energy distributions close to the smeared double step predicted at the QPC output and, consequently, that energy exchanges are small \cite{altimiras2009nesiqhr}.

\begin{figure}[!ht]
\includegraphics[width=\columnwidth]{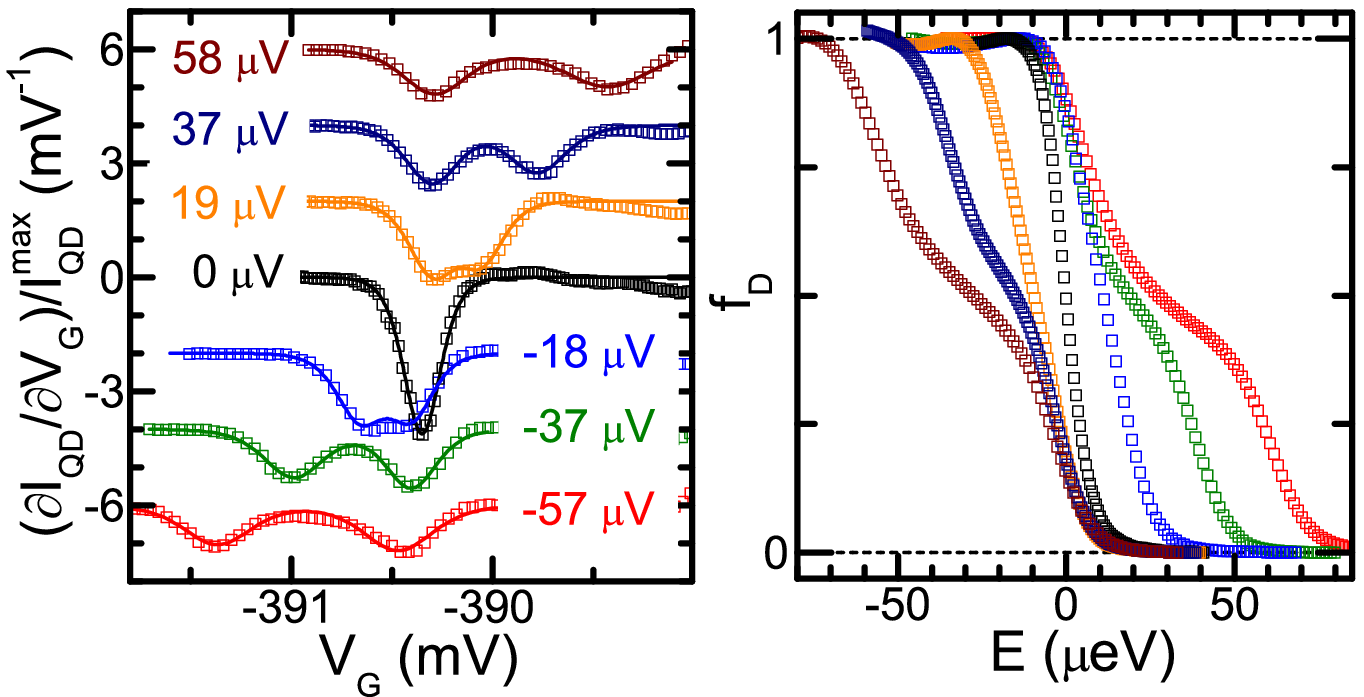}
\caption{The outer EC is driven out-of-equilibrium by biasing at the displayed $\delta V_D$ a QPC set to $G_{\mathrm{QPC}} \simeq 0.5 e^2/h$. Raw data $\partial I_{\mathrm{QD}}/\partial V_{G}$ are proportional to $\partial f_{D,S}/\partial E(E=E_0-e\eta_GV_G)$ measured after a propagation length set to $L=0.8~\mu$m. Left panel: measured sweeps of $\partial I_{\mathrm{QD}}/\partial V_{G}(V_G)$ along the QD-drain signal (symbols) are shifted vertically and aligned. Continuous lines are calculation assuming smeared double steps $f_D(E)$. Right panel: $f_D$ obtained by integrating with $V_G$ the data in left panel, using $\eta_G=0.059$.}
\label{SIFig08um}
\end{figure}

\section{Supplementary discussion}

\subsection{Additional discussion to ascertain the observed energy exchanges are not an experimental artifact}

Figures~2(b) \& 3(b) reveal most directly the presence of energy exchanges along the propagation path. Several possible artifacts giving rise to an apparent energy relaxation are considered here and ruled out.

First, one can ask whether the observed evolution with the propagation length could be attributed to the different QPCs used. We establish it is not the case through three different arguments: \textit{(i)} Experimental configurations $L\in\{2.2,10,30\}~\mu$m use the same voltage biased QPC but the signal is different. \textit{(ii)} Similar hot Fermi dips (not shown) as those at $L=\{10,30\}~\mu$m were obtained at $L=\{12,32\}~\mu$m with a different QPC (light blue metallic gates in Figure~1, also used for $L=4~\mu$m). (iii) These results have been reproduced in three different cooldowns (data not shown), with renewed QD and QPCs.

Second, one can wonder if, somehow, the observed evolution of the signal toward a broad dip as the propagation length is increased could be related to a loss in energy resolution of the QD detector. We checked that it is not the case by measuring the source peak, which remained narrow and was essentially independent of $L$. This implies that the measured source temperature, and consequently the QD energy resolution, was unchanged.

Third, we have checked that these observations do \textit{not} result from charge transfers between ECs as detailed in `Methods' above.

\subsection{Comparison with a recent experiment probing energy currents at $\nu_L=1$}

In a recent experiment, the energy current direction and also the cooling of edge states were investigated at filling factor $\nu_L=1$ \cite{granger2009och}.

In this experiment, the heating of an edge channel is probed qualitatively after various propagation distances from the thermoelectric voltage that develops across a constriction. It was observed that the thermoelectric voltage goes to zero as the propagation distance increases, on $20~\mu$m length scales at $0.1~$K \cite{granger2009och}, which suggests the heated edge states cool down to cold thermal equilibrium.

This energy relaxation toward cold equilibrium is \textit{not} seen here at $\nu_L=2$ on similar length scales. How to explain this seemingly different behavior?

A first possible explanation is that the vanishing thermoelectric signal does not imply unambiguously that the edge states are cooling down to cold equilibrium. Indeed, the thermoelectric voltage could be mostly sensitive to the highest energy electronic excitations (e.g. if the transmission across the constriction has a weak energy dependence). In that case, an energy relaxation toward a hot Fermi function, as observed in the present work, could explain the reduced signal (L.\ Glazman, private communication).

A second explanation, assuming a true relaxation toward cold equilibrium at  $\nu_L=1$, is that the difference with  $\nu_L=2$ is due to the presence at $\nu_L=1$ of low energy spin excitations in the bulk, which was established by the observed fragile spin polarization \cite{[{See }][{ and references therein.}]plochocka_optabsnu1_2009}.

\subsection{Expected limit value of $T_\mathrm{exc}$ for two interacting 1DCFs}

The same minimum allowed value of $T_\mathrm{exc}$ in the EC driven out-of-equilibrium, $T_\mathrm{exc}^{\mathrm{min}}=T_\mathrm{exc}^\mathrm{qpc}/\sqrt{2}$, was obtained within the seemingly different bosonization \cite{degiovanni_plasmon_2010} and 1DCF frameworks \cite{lunde_interaction_2010}.
We here briefly show that this limit value corresponds to temperature equilibration among two parallel 1DCF modes of arbitrary density of states.

Assuming only two 1DCF modes, power balance implies that the injected power at the QPC, which reads
\begin{equation}
P=\tau (1-\tau) \frac{(e\delta V_D)^2}{2h},
\end{equation}
is redistributed among the excess heat currents that are carried along the inner (in) and outer (out) 1DCF channels:
\begin{equation}
P=J_\mathrm{exc}^\mathrm{in}+J_\mathrm{exc}^\mathrm{out},
\end{equation}
with
\begin{equation}
J_\mathrm{exc}^\mathrm{in,out}=\frac{\pi^2}{6h}(k_B T_\mathrm{exc}^\mathrm{in,out})^2,
\end{equation}
where the excess temperatures $T_\mathrm{exc}^\mathrm{in,out}$ depend on the distance to the injection QPC but not the overall edge energy current $J_\mathrm{exc}^\mathrm{in}+J_\mathrm{exc}^\mathrm{out}$.
It is noteworthy that the heat currents above do not depend on the density of states and on the drift velocities of the inner and outer 1DCF channels. This results from the robust velocity - density of states cancelation $v\nu=1/h$ in 1D. Consequently, assuming that the temperatures in the inner and outer 1DCF channels equilibrate at long propagation distances, and taking into account the two pairs of outgoing inner and outer channels, one finds:
\begin{equation}
T_\mathrm{exc} \rightarrow \sqrt{\frac{3 \tau (1-\tau)}{2} \left( \frac{e \delta V_D}{\pi k_B}\right)^2 }=T_\mathrm{exc}^\mathrm{qpc}/\sqrt{2}.\label{SITlimit}
\end{equation}

It is worth noting that, in presence of an additional, third, 1DCF mode, the limit value of the excess temperature would be $T_\mathrm{exc}^\mathrm{qpc}/\sqrt{3}$, about 18\% smaller than the above prediction for two 1DCF edge modes and within error bars from our observations found about 13\% smaller than the limit value in Supplementary Eq.~\ref{SITlimit}.

\begin{table}
\center
\begin{tabular}{|c|c|c|c|c|c|}
  \hline
Ref. & $v_D$ & $n$ & $\mu$ & $d$  \\
 & ($10^5~$m/s) & ($10^{15} \mathrm{m}^{-2}$) & ($\mathrm{m}^2\mathrm{V}^{-1}\mathrm{s}^{-1}$) & (nm) \\
\hline
\cite{zhitenev1994vd} & 0.85 & 1.9 & 70 & 120 \\
\cite{zhitenev1994vd} & 0.55 & 2.3 & 50 & 90 \\
\cite{ernst1997vd} & 1 & 1 & 75 & 130 \\
\cite{talyanskii1993vd} & [1,3] & 1.2 & 100 &  \\
\cite{kamata_vdQHR_2010} & [2.8,4.3] & 3.2 & 170 & 110 \\
\cite{neder2006unexbehamzi} & $>1$ & $\sim1.5$ &  & 85 \\
\hline
\end{tabular}
\caption{Drift velocities at $\nu_L=2$ with metal gates. The main sample parameters are given when known ($n$: electron density, $\mu$: mobility, $d$: depth of 2DEG). Samples in \cite{zhitenev1994vd,ernst1997vd} are fully covered by a metallic gate at the surface, others have metallic side gates. The range of values in \cite{talyanskii1993vd,kamata_vdQHR_2010} was obtained by changing the metal side gate voltage bias (more negative voltages give larger velocities). The data in \cite{neder2006unexbehamzi} permits to obtain a lower bound for $v_D$ from the observed phase rigidity up to an energy of at least $30~\mu$eV in an electronic Mach-Zehnder with an extra length of $2.4~\mu$m along a metal side gate in one of the interferometer's two paths (using Eq.~2 in \cite{neder2006unexbehamzi} with $\varphi<1~$rad).}
\label{SItabVd}
\end{table}

\subsection{Measured drift velocities}

The drift velocity is not measured in our experiment, yet it plays a crucial role to perform the self-consistent test of the 1DCF representation of edge excitations (see article's last paragraph).

The range of drift velocities $v_D\in [0.5,5]~10^5$~m/s used in the article is obtained from different sources that are recapitulated in Table~III. We focus on GaAs/Ga(Al)As devices set to display the integer quantum Hall effect at $\nu_L=2$. Furthermore, the above range of $v_D$ concerns only devices that are either fully covered by surface metal gates or with edges defined by voltage biased metal gates. Note that similar devices without metal gates have a drift velocity typically one order of magnitude larger \cite{[{See e.g. }][{ and references therein.}]kamata_vdQHR_2010}, which would result in an even more stringent failure of the self consistent test described in the article's last paragraph.

%